\documentclass[preprint2]{aastex}
\usepackage{natbib}

\usepackage{epstopdf}

\usepackage[usenames]{color}
\definecolor{MyRed}{rgb}{0.9,0.0,0.0}
\definecolor{MyLightRed}{rgb}{1.0,0.0,0.0}
\definecolor{MyPink}{rgb}{1.0,0.08,0.45}
\definecolor{MyDarkBlue}{rgb}{0,0.08,0.45}
\definecolor{MyDarkGreen}{rgb}{0,0.5,0.0}
\newcommand{\sven}[1]{#1}

\shorttitle{The Low Solar Chromosphere}
\shortauthors{W{\"o}ger, Wedemeyer--B{\"o}hm, Uitenbroek and Rimmele}

\begin{document}

\title{Morphology and Dynamics of the Low Solar Chromosphere}

\author{F. W{\"o}ger}
\affil{National Solar Observatory, P.O. Box 62, Sunspot, NM 88349, USA}
\email{fwoeger@nso.edu}

\author{S. Wedemeyer--B{\"o}hm\altaffilmark{1}}
\affil{Institute of Theoretical Astrophysics, University of Oslo, Postboks 1029 Blindern, N-0315 Oslo, Norway}

\author{H. Uitenbroek}
\affil{National Solar Observatory, P.O. Box 62, Sunspot, NM 88349, USA}

\and

\author{T. R. Rimmele}
\affil{National Solar Observatory, P.O. Box 62, Sunspot, NM 88349, USA}

\altaffiltext{1}{Intra-European Marie Curie Fellow}

\begin{abstract}
The Interferometric Bidimensional Spectrometer (IBIS) installed at the Dunn Solar Telescope of the NSO/SP is used to investigate the morphology and dynamics of the lower chromosphere and the virtually non-magnetic fluctosphere below.
The study addresses in particular the structure of magnetic elements that extend into these layers.
We choose different quiet Sun regions in and outside coronal holes.
In inter-network regions with no significant magnetic flux contributions above the detection limit of IBIS, we find intensity structures with the characteristics of a shock wave pattern.
The magnetic flux elements in the network are long lived and seem to resemble the spatially extended counterparts to the underlying photospheric magnetic elements.
We suggest a modification to common methods to derive the line-of-sight magnetic field strength and explain some of the difficulties in deriving the magnetic field vector from observations of the fluctosphere.
\end{abstract}



\section{Introduction}
\begin{figure*}[t]
\centering
\includegraphics[width=\textwidth]{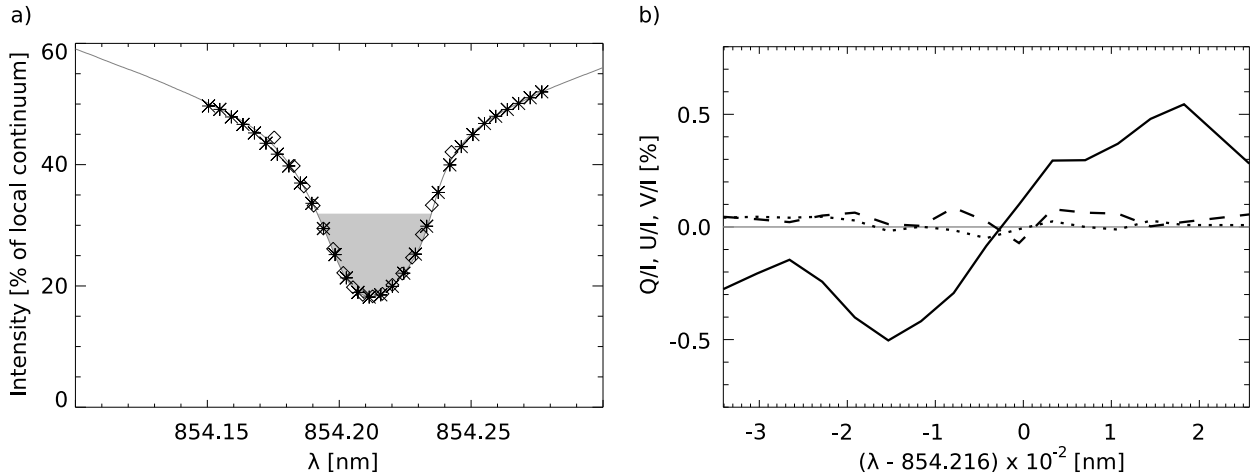}
\caption{a) The Ca II infrared line (854.2\,nm). Displayed as $*$ are the sampled wavelength points (30) of data set 2, the $\Diamond$ indicate those of data set 1. The shaded area indicates the wavelength and intensity regime within which bisectors for the use with the hybrid bisector-COG were calculated (see text).
b) Stokes profiles of the strong magnetic features located at [11,16]\,arcs in Fig.~\ref{fig:mag}~c) (black solid: Stokes V/I; dashed: Stokes Q/I; dotted: Stokes U/I).}
\label{fig:wavepoints}
\end{figure*}
Understanding the chromosphere and its magnetic structure is important to progress in understanding the solar atmosphere as a coupled phenomenon from the photosphere to the corona.
A comprehensive physical model of the chromosphere must be capable of explaining a multitude of observations including, e.g., Calcium K grains \citep{1968SoPh....3..367B}, the magnetic canopy \citep{1983SoPh...87...37J}, and chromospheric oscillations \citep{1993A&A...274..584K,1993ApJ...414..345L}.
Constructing such a model is complicated by many things, and detailed observations such as those presented in this work are important to constraint this effort.
The few available chromospheric diagnostics are unfortunately subject to complicated formation processes, for which -- in most cases -- the simplifying assumption of local thermodynamic equilibrium cannot be made, causing severe difficulties for the quantitative comparison between models and observations.
Furthermore, fluctuations on small spatial and temporal scales are very challenging for both modeling and observation.
Advances in both fields has lately led to an increase of attention paid to the solar chromosphere, sparking a controversy about its ``true'' physical nature.
A recent summary of this controversy has been given by \citet{2009A&A...494..269V}.

\begin{figure*}[t]
\centering
\includegraphics[width=\textwidth]{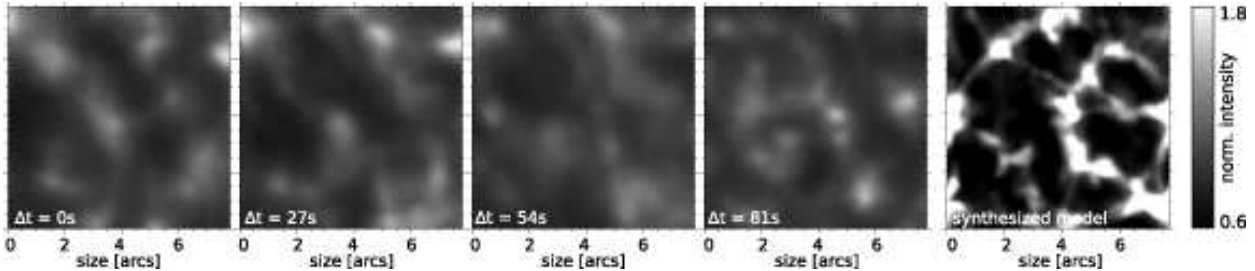}
\caption{Sequence of subframes of a weak-field region in the Ca II infrared line core, in comparison to the synthesized model that was degraded with the point spread function of the Dunn Solar Telescope (right panel). Intensity is scaled to the same range in all panels.
}
\label{fig:corr}
\end{figure*}
Recent high-resolution observations suggest that in the quiet Sun there exists a weak-field domain below the classical canopy \citep{2006A&A...459L...9W}.
The prominent magnetic fields of the canopy are major constituents of the chromosphere in the stricter sense as it is seen in the H$\alpha$ line.
The weak-field domain below is referred to as ``fluctosphere'' hereafter -- a term first used by \citet{2008IAUS..247...66W} 
\sven{and previously named ``clapotisphere'' \citep{1991SoPh..134...15R, 1995ESASP.376a.151R}.}
The fluctosphere in numerical radiation hydrodynamic models is generated by interference of (acoustic) shock waves, which are excited in the photosphere and propagate upwards into the layer above \citep[cf.][]{1994chdy.conf...47C}.

At heights around 1000\,km above $\tau=1$ this results in an apparent temperature pattern that has similarity to reversed granulation, yet changes with typical timescales between 20--30\,s \citep{2004A&A...414.1121W}.
This pattern is the result of the steep temperature increase at the vertices of the interfering wavefronts, whereas the enclosed post-shock regions are cool on average.

The recent ability to synthesize spectra from models depicting the fluctosphere \citep{2009SSRv..144..317W} allow their direct comparison to observations of this atmospheric region, under both morphological as well as dynamical aspects. Acoustic shocks have been analyzed in detail by \citet{2009A&A...494..269V}, who use the Ca II infrared line at 854.2\,nm as diagnostic of what they call the ``mid-chromospheric'' region of the solar atmosphere.
It has been shown by \citet{2008A&A...480..515C} that the Ca II infrared triplet provides a convenient and accessible diagnostic of the chromosphere.
As \citet{2001ApJ...554..424J}, \citet{2009A&A...494..269V} notice a strong influence of magnetic field on acoustic processes within that regime of the solar atmosphere, and conclude that this influence may be larger than generally expected.
Nevertheless, they also confirm the existence of regions with shock behavior that is
similar to what is predicted by numerical models with no or at least weak magnetic field only.

A fundamental knowledge of the chromospheric magnetic field topology and its dynamics thus seems to be the next unavoidable step towards a comprehensive understanding of the solar atmosphere.
However, the detailed magnetic properties of the ``mid-chromosphere'' are widely unknown so far because of the obstacles that need to be overcome in observational techniques.
This in particular applies to the weak-field regions in the quiet Sun.
It has been suggested that magneto-acoustic heating plays a role in the chromospheric heating \citep[e.g.][]{2005ApJ...631.1270H, 2008ApJ...680.1542H}, even though the predominant mechanism is still debated.

Here we present and interpret for the first time high spatially resolved spectro-polarimetric measurements in the fluctosphere above strong photospheric magnetic features.
We will analyze the fine-structure of the line wing and core intensity in the Ca~II infrared line in comparison to synthetic maps from numerical models.
The paper is structured as follows.
In Sect.~\ref{sec:data}, we present the data observed along with the methods used for calibration.
Section~\ref{sec:top} describes the algorithm and result of the analysis of the three-dimensional topology inherent to a magnetic structure found in our data, whereas in Sect.~\ref{sec:seq} we show the results of our time sequence analysis.
We conclude with a summary of our results in Sect.~\ref{sec:con}.

\section{Data}\label{sec:data}

\subsection{Data sets}

\paragraph{Data set 1:} We observed a quiet Sun region located in a coronal hole at disc-center on May 26, 2007 using the Ca\,II infrared line at 854.2\,nm.
Several persistant G-band bright points are visible in the field of view, forming a network element.
The overall observation lasted for 26.5 minutes.

The IBIS narrowband channel was set up in spectro-polarimetric mode to scan the line core of the Ca II line in two spatial dimensions using six modulation states and an exposure time of 35\,ms.
The line was scanned with 17 wavelength steps using a step width in wavelength of 4.3\,pm with a full width at half maximum (FWHM) transmission of 4.6\,pm (see Fig.~\ref{fig:wavepoints}).
This procedure resulted in a overall cadence of approximately 27\,s for one data cube containing two spatial dimensions and a third spectral dimension.

\begin{figure*}[t]
\centering
\includegraphics[width=0.8\textwidth]{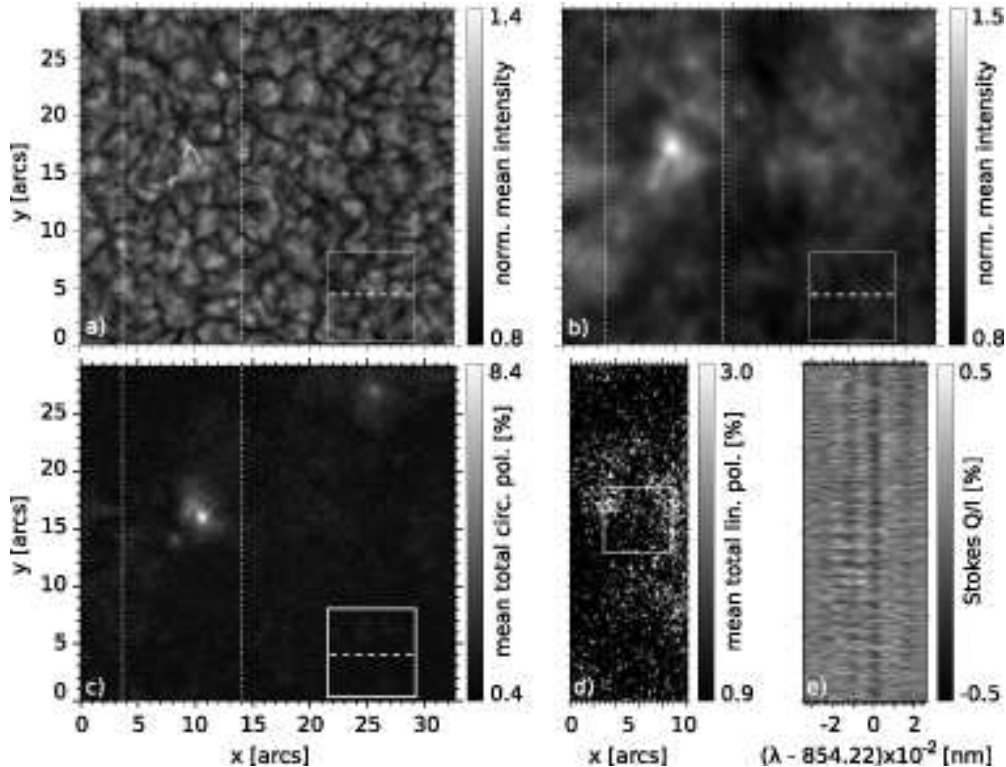}
\caption{Data set~1: Temporal averages over 26.5\,min of
a) G-band intensity,
b) Ca II line core intensity,
c) total circular polarisation.
The inscribed solid box in a) -- c) indicates the location of the subfield presented in Fig.~\ref{fig:corr}, the dashed line in therein indicates the position of the simulated slit in Fig.~\ref{fig:spectra}.
d) Total linear polarisation of the right half of the field of view indicated with the dotted box in a) -- c).
e) Stokes Q/I profiles of enhanced total linear polarization ($>$ 1.9\%) in the square inscribed in d).
The average of the displayed Stokes Q/I profiles leads to the Stokes Q/I profile visible in Fig.~\ref{fig:wavepoints}.
}
\label{fig:mag}
\end{figure*}

A subregion of the data set displayed for the Ca II infrared line core is shown in Fig.~\ref{fig:corr} for a small part of the time sequence, Fig.~\ref{fig:mag} gives an overview over the whole field of view, where total circular polarization was defined as
\begin{equation}
T_{c} =  \int_{\lambda_{a}}^{\lambda_{b}} \, \mbox{d}\lambda \sqrt { (V(\lambda)/I(\lambda))^2 }
\end{equation}
and total linear polarization was defined as
\begin{equation}
T_{c} =  \int_{\lambda_{a}}^{\lambda_{b}} \, \mbox{d}\lambda \sqrt { (Q^2(\lambda)+U^2(\lambda))/I^2(\lambda) }.
\end{equation}
This data set was observed to acquire further insight into the dynamics of the chromosphere at locations of strong and weak magnetic flux.
Unfortunately, after reduction of the spectro-polarimatric data of data set 1, the Stokes U component has been found to contain fringes.
The fringes were compensated, however, at the cost of a reduced signal to noise ratio.
In addition, the Stokes Q component was plagued by a weak reflection in the field of view, which hampered the Stokes Q signal in the right half of the FOV.
Overall, the Stokes Q and U components were barely above the noise level, even after spatial and temporal averaging.

\paragraph{Data set 2:}
Complementary to data within a coronal hole, we have observed the Sun on disk-center outside of a coronal hole.
These observations were carried out on September 22, 2008.
In contrast to the coronal hole observations, no time sequence was acquired.
However, the exposure time was increased to 90\,ms for enhanced sensitivity.
In addition to the Ca~II infrared spectra, the Fe~I line at 630.2\,nm
was observed with IBIS subsequently with a time difference of 1 minute.
For these observations, the number of wavelength points was increased to 30 in the Ca II infrared line to cover a larger part of the spectrum (Fig.~\ref{fig:wavepoints}).
The Fe I line was scanned with 30 wavelength points with a stepwidth of 3.2\,pm.
The observations thus includes both the 630.1 and 630.2 lines, as well as the telluric blend spearating these two lines.
The FWHM transmission of IBIS at this wavelength is about 2.3\,pm.

This data set was acquired complementary to data set~1 to gather additional information, which is used to check the magnetic field information retrieved from the Ca II infrared line.

\subsection{Data calibration}

The IBIS instrument consists of two channels, a narrowband channel with two tunable Fabry-Perot interferometers and a broadband channel for reference \citep{2006SoPh..236..415C}.
The broadband images of IBIS were calibrated with the standard procedure of dark subtraction and gain table application and reconstructed using the speckle image reconstruction algorithm modified for high-order adaptive optics corrected data detailed in \citet{2008A&A...488..375W}.

The IBIS narrowband data were calibrated using the following procedure.
In a first step, the gathered flat and dark images were averaged.
Because the Fabry-Perot interferometers of IBIS are located in a collimated beam, each pixel in a single narrowband exposure is affected by a blueshift \citep{2006SoPh..236..415C}.
The magnitude of the blueshift was estimated from the dark corrected data cube of average flat images by determining for each spatial pixel the line core position to sub-pixel accuracy.
The resultant two-dimensional blueshift map then fitted with its theoretical form, a paraboloid.
In a subsequent step, the mean line profile was computed by averaging the blueshift corrected single profiles.
To compute the gain table, each line profile at a pixel position in the data cube was
divided by the mean profile which was shifted to match the line core position at that specific pixel.
After the application of the gain table to the data, the narrowband images of IBIS were destretched using the broadband images in combination with their reconstruction.
Finally, all line profiles were blueshift corrected, leading to a pixel artefact free data cube, which served as input for the Stokes parameter calibration.

The polarization calibration is accomplished using procedures originally developed for the Advanced Stokes Polarimeter (ASP) \citep[e.g.][]{1993ApJ...418..928L}, that were adapted for usage with IBIS spectro-polarimetric data.
The procedure is wavelength independent, and has been published for IBIS in detail by \citet{2009ApJ...700L.145V}.
The polarization calibration includes removal of residual crosstalk between the Stokes parameters.
Recently, \citet{judge} has compared IBIS spectro-polarimetric data simulatenously observed to that of the spectrograph (SOT/SP) onboard the HINODE satellite, the result of which indicates that the calibration processes yield very similar results.
A publication describing in detail the complete calibration of spectroscopic and spetro-polarimetric data acquired with IBIS is currently in preparation \citep{tritschler}.

\begin{figure*}[hpt]
\centering
\includegraphics[width=\textwidth]{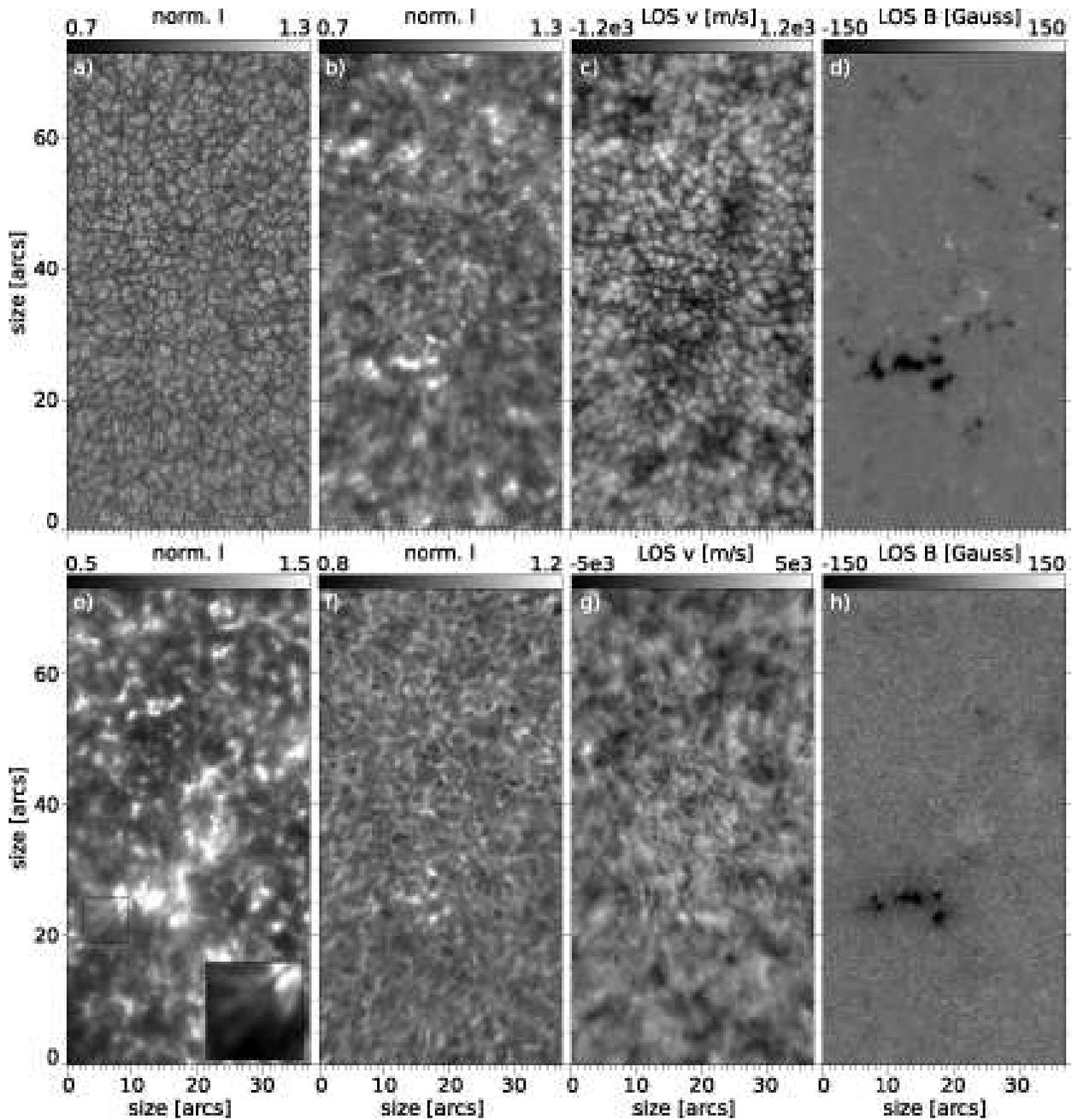}
\caption{
Data set 2, acquired on September 22, 2008. a) IBIS continuum image, b) Fe I (630.2\,nm) core image, c) line-of-sight velocity of Fe I d) line-of-sight magnetic flux density in Fe I, e) Ca II (854.2\,nm) core intensity, with a contrast enhanced close-up of the marked fibrils in the subpanel, f) Ca II wing intensity (+ 0.073\,nm), g) line-of-sight velocity of Ca II and h) line-of-sight magnetic flux density in Ca II. Velocities and magnetic flux densities have been calculated using the genuine COG method analyzed in detail in \citet{2003ApJ...592.1225U}.}
\label{fig:scandata}
\end{figure*}

\section{Topology of magnetic network elements}\label{sec:top}
For data set~2, we derive the topology of strong magnetic fields reaching from the photosphere into the chromosphere.
An example of the data quality is given in Fig.~\ref{fig:scandata}.
A strong magnetic feature consisting of photospheric bright points can be seen in the intensity map at a continuum wavelength of 828.04\,nm.
The chromospheric counterpart is highly structured and consists of many fibrils.
Using the method put forward by \citet{1979A&A....74....1R}, and later also \citet{2003ApJ...592.1225U}, we derive the LOS velocity and the magnetic flux density.
The results for the 3D topology of the magnetic element are very similar for the Fe\,I and the Ca\,II\,IR line (see Fig.~\ref{fig:scandata}, panels d) and h)).
It appears that both predominantly map the magnetic flux in the photosphere.
This is due to the fact that the result of the COG method represents an average over the entire Ca II IR line profile which encodes the chromospere only within a very small wavelength interval around the line core, whereas the line wings -- that encode mostly the photosphere -- make out the major part of the profile.
Thus, the result is significantly biased towards lower atmospheric regions.

\subsection{The hybrid bisector-COG method}
The center of gravity (COG) method \citep{2003ApJ...592.1225U} to derive LOS magnetic flux is very robust, but has the disadvantage that height fluctuations in the field cannot be recovered.
For such an analysis, usually a bisector analysis is employed.
The latter is unfortunately sensitive to noise.
In order to increase the reliablility of our topology analysis, we devise a hybrid method combining the bisector method with the COG method.

In the algorithm, the following steps are performed:
\begin{enumerate}
\item Compute the wavelengths at which the red and blue line wing of both the \mbox{I+V} and \mbox{I-V} spectra have the same intensity. This can be pictured as finding the two points ($\lambda_{a}$ and $\lambda_{b}$) of intersection of the line profile with a horizontal line.
\item Use the two acquired wavelength ranges to compute the COG of the I+V and I-V spectra with
\begin{equation}
\lambda_{+} = \frac
{\int_{\lambda_{a}}^{\lambda_{b}}\mbox{d}\lambda\; \lambda\; \{\mbox{(I+V)}(\lambda_{a}) - \mbox{(I+V)}(\lambda)\}}
{\int_{\lambda_{a}}^{\lambda_{b}}\mbox{d}\lambda\; \{\mbox{(I+V)}(\lambda_{a}) - \mbox{(I+V)}(\lambda)\}}
\end{equation}
for \mbox{I+V} and analogously $\lambda_{-}$ for \mbox{I-V}.

\item The value for the magnetic flux density along the line-of-sight is then computed from the difference of the COG positions of the \mbox{I+V} and \mbox{I-V} profiles, using
\begin{equation}
\mbox{B}_{\mbox{\tiny LOS}} = \frac{\lambda_{+}-\lambda_{-}}{2} \  \frac{4\ \pi\ m\ c}{e\ g_{L}\ \lambda_{0}^{2}},
\end{equation}
where $\lambda_{0}$ is the central wavelength of the line, $\lambda_{\pm}$ are the wavelength positions of the centroids of I$\pm$V, $g_{L}$ is the effective Land{\'e} factor, and $e$ and $m$ are the electron charge and mass, respectively, in SI units \citep[see][and references therein]{2003ApJ...592.1225U}.
\end{enumerate}
Step 1 is needed for two reasons:
First, the selection of many different intensity values allows to analyze the magnetic flux density at different formation height ranges, which is also true for the genuine bisector method.
Secondly, an uneven cut of the \mbox{I+V} and \mbox{I-V} would bias the measurement of the COG value.
However, it is obvious that this method only allows an analysis of the magnetic flux density in non-emitting lines, as an emission peak would make Step~1 and Step~2 very difficult.
An emission peak would create four instead of two points of intersection and also bias the computation of the COG.
Another weakness of the described method is sensitivity to asymmetries in the Stokes V profiles.
Such asymmetries are found e.g. in a detailed analysis of a Quiet Sun region near network by \citet{2007ApJ...670..885P} on a common basis, and can lead to biased measurements of magnetic flux.
In what follows, we test the proposed method with synthetic profiles that inhibit similar behavior.

\subsection{Check of method}
We have verified the feasibility of this approach by applying it to a magnetohydrostatic model of a funnel-like expanding fluxtube.
It originates in the deep photosphere with a footpoint of 500\,km in diameter and has a magnetic field strength of 2500\,Gauss.
NLTE spectra for all four Stokes parameters were subsequently synthesized for the Ca~II~IR line.
\begin{figure*}[t]
\centering
\includegraphics[width=\textwidth]{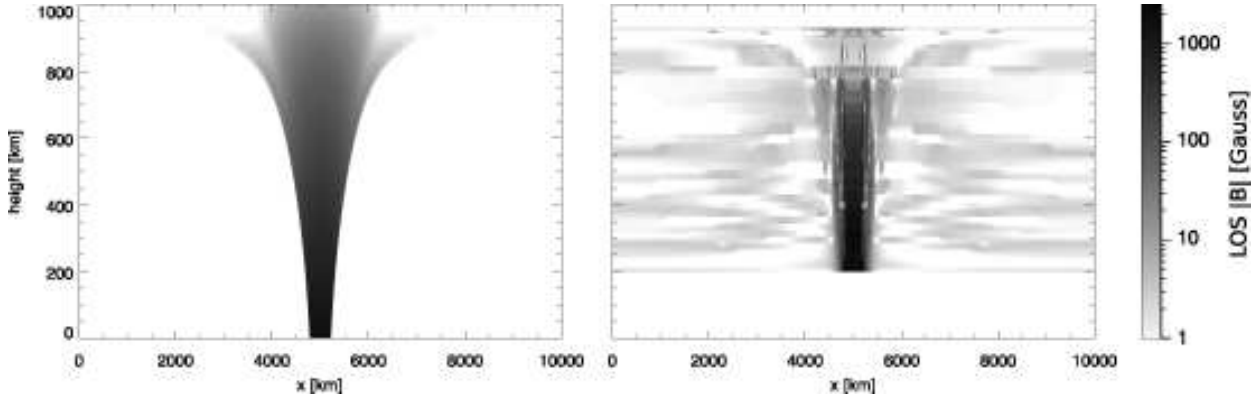}
\caption{Result of the accuracy test of the proposed hybrid bisector-COG method. a) input model, and b) recovered flux density. The height scale in b) was derived using calculations of $\tau_{\lambda}=1$. The gray scale encodes the magnetic flux density and is the same for both panels.}
\label{fig:methodtest}
\end{figure*}
The result is shown in Fig.~\ref{fig:methodtest}.
The absolute magnetic flux density was accurately recovered where magnetic field was present, as was the photospheric topology of the fluxtube.
No significant jumps occur between the results computed from adjacent intensity levels.
This is likely due to the COG algorithm that reduces noise in the measurement by including many more points in the derivation of the magnetic flux density than the simple bisector method.
Interpretation of the three-dimensional topology in the higher layers becomes difficult due to the fact that a bisector does not necessarily correspond to a thin atmospheric layer.

The effective formation height was derived for the Ca II infrared line by averaging the geometrical heights of the locations where $\tau_{\lambda}=1$ over each of the wavelength intervals computed in Step~1.
In general, the formation height increases from line wing towards the line core.
In addition, the ``Wilson effect'' causes the height of $\tau_{\lambda}=1$ to differ
in magnetic and non-magnetic regions.
The strong magnetic flux causes the opacity within the flux tube to be lower compared to the
ambient medium, resulting in contributions to line core signal from layers at about 200\,km lower
compared to the $\tau_{\lambda}=1$ height in the non-magnetic surrounding.
The Wilson effect has been compensated in Fig.~\ref{fig:methodtest}~b), which demonstrates the well recovered topology of the flux tube.

The comparison of the input model in panel~a) and the reconstruction in panel~b) of \ref{fig:methodtest} serves now as reliability test and error estimate for the proposed method.

The recoverable height interval is here restricted by the line properties such as formation height and the chosen wavelength intervals in Step~1.
Qualitatively, the topology of the modeled fluxtube was recovered.
The regions without magnetic field in the input model show a weak signal in the reconstruction corresponding to field strengths of up to $(6 \pm 1)$\,G.
These artefacts can primarily be attributed to the fact that the Stokes V spectra exhibit features like spikes when discontinuities occur along the LOS, biasing the I+V and I-V spectra and introducing a fake magnetic signal.

\subsection{Reconstructed field topology in data set~2}
We now analyze dataset 2 and use 101 intensity values ranging from 18.5\% (Ca II IR line core) to 32\% of the local continuum intensity (Fig.~\ref{fig:wavepoints}).
We used a median filter with 5$\times$5 pixels (corresponding to 0.85$\times$0.85\,arcs$^2$) to reduce salt and pepper noise introduced by the measured spectra.

\begin{figure*}[t]
\centering
\includegraphics[width=\textwidth]{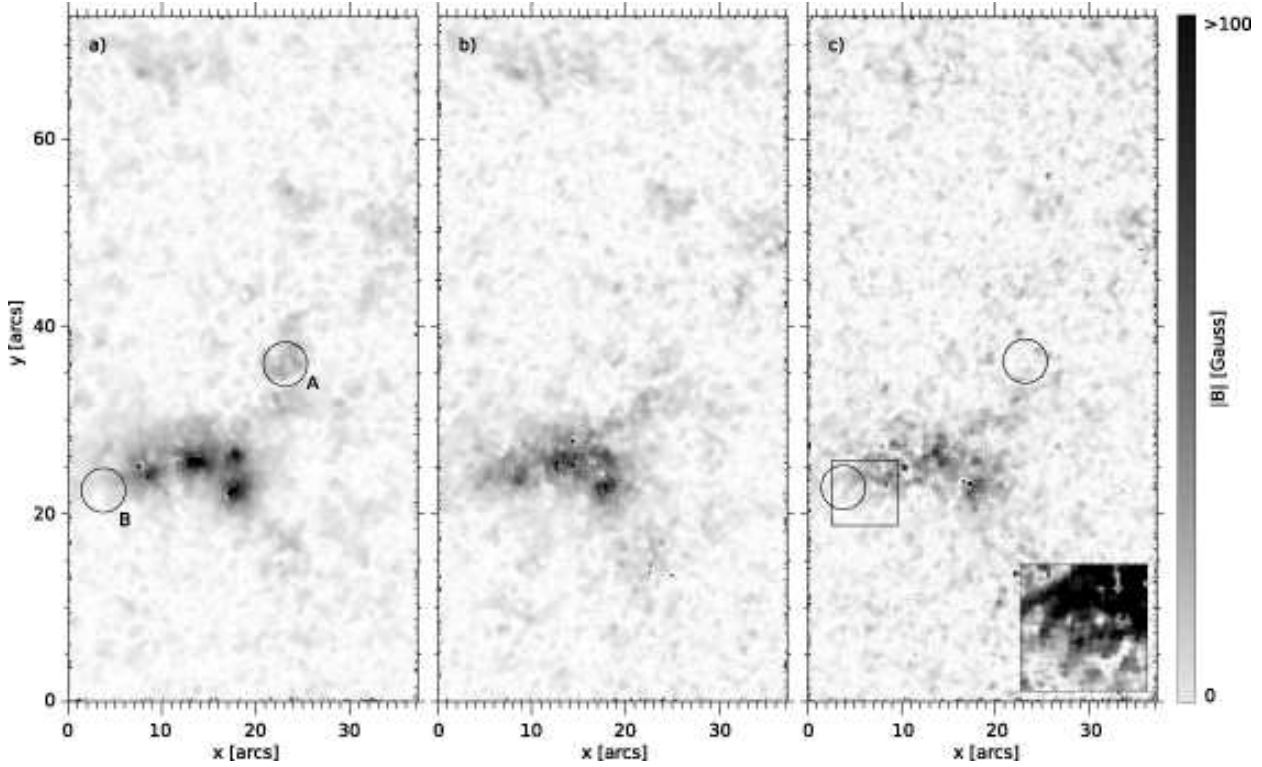}
\caption{Cuts through the recovered 3D topology of the magnetic flux density by applying the hybrid bisector-COG method to the Ca II IR line scan of dataset 2.
Panel a) is at 32\% of the local continuum intensity (photospheric) (see Fig.~\ref{fig:wavepoints}), b) at 20.2\% and c) at 18.5\% (fluctospheric), with a contrast enhanced close-up of the fibris marked by the box in a subpanel.
At location~A, a patch of magnetic flux along the line-of-sight seems to disappear with increasing height (also see subpanel).
Location~B is an example for the appearance of a filament like structure with height.
}
\label{fig:3dtop}
\end{figure*}
First we apply the COG method to the Fe\,I line data and obtain a magnetic flux density of about 350\,Gauss for the strong feature in the FOV, likely because the small-scale magnetic elements have not been resolved, yielding a decreased field strength because the signal is averaged over the area of a pixel.
The COG method is sufficient here because this spectral line is formed over a comparatively small height range in the atmosphere.
The formation of the Ca\,II IR line, on the other hand, extends over more
than a 1000\,km.

Nevertheless, the computed magnetic flux density can serve as a well-tested value for those retrieved using the Ca II IR line.
Indeed, the result fits well with the 250\,Gauss obtained from the measured Ca II IR line using the same method, which started at $\sim$50\% of continuum intensity; the formation height of the magnetic signal of the Ca II IR line expectedly appears to be higher than that of Fe I.
The value drops further to 125\,Gauss with our hybrid-COG method when using bisectors starting at 32\% of the local continuum intensity.
This is not surprising because we have restricted ourselves even further to the line core.
The result of the hybrid bisector-COG analysis of our data can be seen in Fig.~\ref{fig:3dtop}.

It is worth to point out two situations visible in the data.
A patch of magnetic flux directed in the direction of the observer seems to disappear when restricting oneself to the line core with the derivation (Fig.~\ref{fig:3dtop}, location A), while at the same time, a filament like structure appears (Fig.~\ref{fig:3dtop}, location B).
Obviously, the patch of location A does not reach far up into the solar atmosphere, likely connecting to regions with opposite polarity in the photosphere.
The filament like structure, on the other hand, appears in a location of enhanced core intensity of the Ca II IR line (see also Fig.~\ref{fig:scandata}).
To our knowledge, this is the first time a filament-like structure has been detected in magnetic signal.

The confined photospheric magnetic structure appears to become fragmented at higher atmospheric levels.
Overall, it appears that the structure of the magnetic flux density along the LOS, i.e., essentially the vertical component, becomes smaller and weaker.
This can be explained with a magnetic funnel expanding with height.
In that case, the magnetic field becomes increasingly horizontal with height, so that the vertical flux component decreases while the horizontal component increases.

Due to instrumental limitations, weak horizontal fields cannot be detected in data set 2.
The failure to measure the horizontal field in total linear polarization can be explained with the finite sensitivity of these particular IBIS observations -- no temporal averaging was possible as data set 2 only consists of a single line scan.

Using temporal averaging it is possible to increase the signal to noise ratio in the spectro-polarimetric data by decreasing photon noise, while losing the ability to perform any dynamic analysis.
To detect horizontal magnetic fields, we have temporally averaged data set 1.
The integrated exposure time of the resulting maps is equivalent to  12.5\,s.
The result is shown in Fig.~\ref{fig:mag}~a)--d).
The Stokes parameters averaged over the time sequence are displayed in Fig.~\ref{fig:wavepoints}~b).
There are strong magnetic features (around x=11\,arcs, y=15\,arcs), which show up as distinct signals in the total circular polarization signal with a diameter of 1\,arcs.
This feature can also be seen at the same location in the photosphere, particularly well in G-band images.
Only the summed up data have sufficient signal to noise ratio to study the change in topology of the magnetic flux density that extends from the photosphere into the chromosphere.
The photospheric footpoints of the magnetic features are advected with the convective motions.
As the duration of the time sequence is rather short, the features roughly remain at the same location.
Thus, in the averaged G-band images the location of the magnetic features are clearly visible (Fig.~\ref{fig:mag}~a)).
Temporal averaging reveals a structure in total linear polarization encirceling the magnetic elements visible in total circular polarization (Fig.~\ref{fig:mag}~d)).
This structure supports the view of a horizontal magnetic flux component that increases from photosphere to chromosphere: as the magnetic flux becomes weaker with increasing height and distance from its footpoint in the photosphere, it becomes unmeasurable even with an equivalent exposure time of 12.5\,s.

\subsection{Discussion}
When analyzing the expected polarimetric signal of e.g. the hydrostatic model of an expanding flux tube described above and displayed in Fig.~\ref{fig:methodtest}~a), a very {\em weak} total linear polarization signal is expected (of the order of 0.03\% of the measured intensity), which is below the sensitivity of IBIS even with 12.5\,s integrated exposure time.
The {\em measured} linear polarization signal is of the order of 0.1\% of the measured intensity (Fig.~\ref{fig:wavepoints}), which is barely above the noise level, whereas amplitude and spatial extent of the input model fluxtube at photospheric heights fit well with the measured total circular polarization signal in our data sets.

Possible explanations for this finding are summarized in what follows.
\begin{itemize}
\item The origin of the measured signal cannot be attributed easily to a certain height layer in the solar atmosphere.
Significant contributions to the polarization signal may originate from atmospheric regions with (a)~strong magnetic field for which the contribution function is small, and (b)~weak magnetic field for which the function has large values.
The total circular polarization signal may therefore originate from a different atmospheric layer than the total linear polarization signal.
These issues may be the reason for the agreement of the total circular polarization signal with the model of an expanding flux tube, as its signature originates from deeper layers than the fluctosphere. In these regions the model may predict the flux tube expansion correctly.
\item The "large" values of the total linear polarization may indicate that
loop-like structures prevail which connect to magnetic elements in the photosphere.
In this case the total linear polarization signal (and thus the horizontal magnetic field) may originate from atmospheric regions that lie well below the fluctosphere in the upper photosphere \citep[e.g][]{2008ApJ...680L..85S,2008ApJ...672.1237L,2009A&A...495..607I}
In those regions, the magnetic field lines are more confined and produce a stronger polarization signal.
\item The model of a static flux tube that expands as a funnel-like structure into the chromosphere likely is too simplified to create an accurate picture of the dynamic chromosphere. Recent observations and simulations suggest that magnetic field elements may have a more complex and entangled structure than thought before \citep[see recent reviews by][and references therein]{2009SSRv..144..317W, 2009A&A...494..269V, 2009arXiv0905.3124P}.
\end{itemize}

\section{Small-scale intensity pattern in inter-network regions}\label{sec:seq}
\begin{figure*}[t]
\centering
\includegraphics[width=\textwidth]{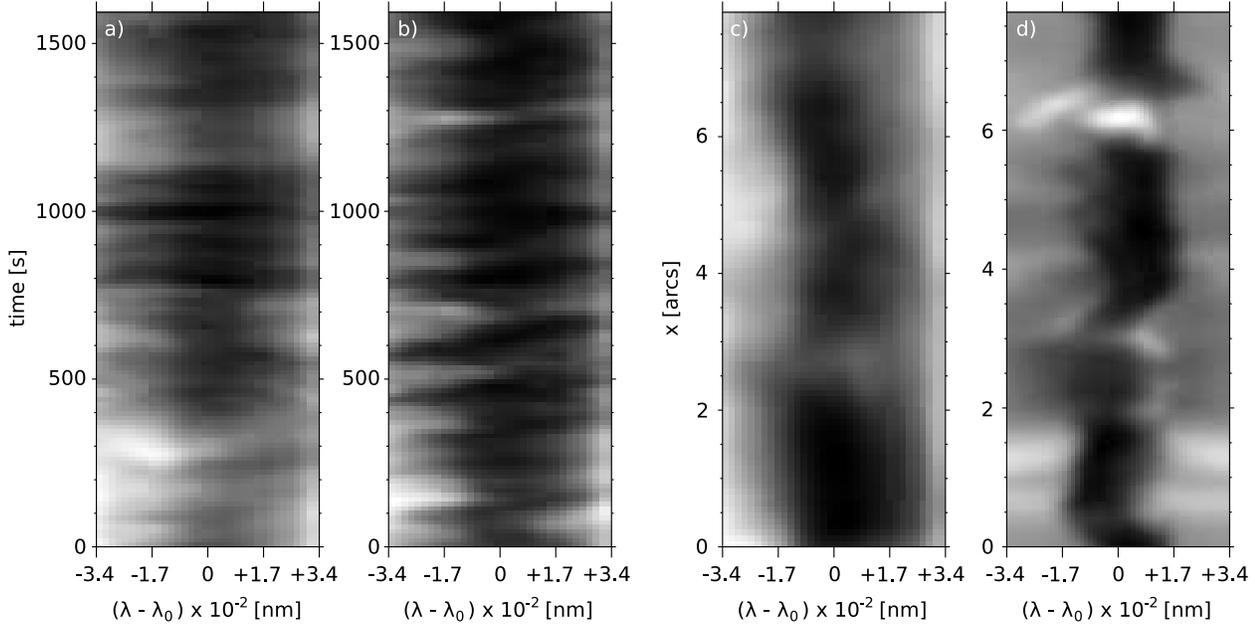}
\caption{a) Time variation of the spectrum of the Ca II IR line at the location [11,16]\,arcs in the field of view of dataset 1 (magnetic element). b) Like panel a), except for a weak-field location at [24,3]\,arcs. c) Spatial spectrum of dataset 1 at t=530\,s along the slit indicated in Fig.~\ref{fig:mag}~b) with a length of $\sim$7.7\,arcs. d) Spatial variation of the synthesized model degraded using the point spread function of the Dunn Solar Telescope along a slit at position $x = 2.75$\,arcs, with a slit length of 7.7\,arcs.}
\label{fig:spectra}
\end{figure*}

The short intensity map sequence displayed in Fig.~\ref{fig:corr} is characterized by very short-lived grain-like features of enhanced intensity.
Their sizes are of the order of an arcsec or less.
While such Ca\,grains are well-known phenomenon, the substructure of the background has been less studied -- due to instrumental limitations.
It is the combination of a very narrow transmitted wavelength range, a short exposure time, and the high spatial resolution due to the well performing adaptive optics system that enable us resolve the pattern between the grains.
We now clearly see dim strands connecting the grains.
While the grains appeared as isolated features before, they now appear to be the nodes of a mesh-like pattern with rather dark regions in-between.
Short-lived ``calcium grains'' have been known for a long time \citep[e.g.][]{1993A&A...274..584K} not only in Ca II K, but also in the Ca II IR line.
The formation of ``calcium grains'' has been explained using 1D simulations by
\citet{1997ApJ...481..500C} as propagating shock waves.
Indeed we find grains in connection with a clear signature of shock waves in the time variation of the spectrum  (Fig.~\ref{fig:spectra}).
This signature is typical throughout the weak-field regions of data set~1.

The intensity pattern at small spatial scales resembles the pattern produced by interaction of propagating shock waves as seen in radiation hydrodynamic simulations of \citet{2004A&A...414.1121W}.
We therefore compare the observed line core maps with the synthetic intensity maps described in \citet{2009SSRv..144..317W}.
The latter were calculated with the non-LTE radiative transfer code MULTI \citep{1986UppOR..33.....C} from a time-dependent radiation hydrodynamics simulation with \mbox{CO$^5$BOLD} \citep{2008asd..soft...36F}.
The line core intensity originates from a height range of around 1000-1500\,km \citep[see the contribution functions by][]{2006ASPC..354..313U}.
The subregion shown in Fig.~\ref{fig:corr} was chosen to match the size of the model.
The synthetic map in the figure was degraded by convolution with a theoretical point spread function (PSF) of the Dunn Solar Telescope.
As described in \citet{2009SSRv..144..317W}, this PSF is rather optimistic.
It is likely that the instrumental straylight is underestimated and that residual atmospheric distortions are still present even though the observations were acquired with the high-order adaptive optics system operating near the diffraction limit, leading to a too high contrast compared to the observations.
And still most of the filigree fine-structure in the original synthetic maps is lost after application of the PSF.
What remains is very similar to the observation:
A mesh-like pattern enclosing dark regions and bright grain-like features at the vertices of the mesh.
In the model, the brightening occurs at locations with high gas temperature as a result of the compression of the gas in the collision zone between neighboring shock waves.

The aforementioned simulations do not take into account magnetic fields and thus represent highly idealized conditions.
The observations displayed in Fig.~\ref{fig:corr} were taken in a very quiet region within a coronal hole and are therefore most likely as quiet as they could possibly be.
Indeed, the line core images exhibit no fine-structure with ``fibrils'' or ``threads'' which are otherwise seen in the vicinity of stronger magnetic fields
\citep[cf. data set 2 and, e.g.,][]{2005A&A...435..327R,2009A&A...494..269V}.
Although no clear pattern is discernible neither in the total circular nor in the total linear polarization measured in the subregion (see Fig.~\ref{fig:mag}), weak magnetic fields at or below the detection limit could well be an integral component of the fluctosphere even under such extreme conditions.
The simulations by \citet{2006ASPC..354..345S}, which include a weak magnetic field in the fluctosphere, still exhibit a shock wave pattern similar to those seen in the field-free simulations.
The simulations by \citet{2008ApJ...679..871M} show a similar pattern, which is also connected to a gas temperature distribution with a hot shock wave component and a cool background.
Already the simulations by \citet{2000ApJ...541..468S} produced a shock-induced pattern can, although the relatively coarse grid spacing makes it harder to discern.

In Fig.~\ref{fig:spectra} we compare the spectra of the non-magnetic model degraded with the point spread function of the Dunn Solar Telescope with one of our measurements.
The figure shows the temporal evolution of the spectra in the magnetic element visible at [11,16]\,arcs, e.g., in Fig.~\ref{fig:mag}~c) and at [24,3]\,arcs, a region with no appearant magnetic signal (Fig.~\ref{fig:spectra}~b).
The spectrum from the region without appearant magnetic flux is almost identical to those visible in \citet{2009A&A...494..269V}, showing frequent and regular shock events.
The spectrum in the magnetic element (Fig.~\ref{fig:spectra}~a) shows a different picture: here, shocks happen much rarer and irregular.
It demonstrates the influence of magnetic fields on the generation and  propagation of shock waves as it has been frequently observed \citep[cf.][]{2009A&A...494..269V}.
In addition, events in the spectrum appear to be mostly simultaneous in blue and red wing, whereas changes in the non-magnetic temporal spectra (Fig.~\ref{fig:spectra}~b) happen with temporal delay indicating acoustic events.

Shock events (e.g., at $t=530$\,s) produce similar structures in synthetic and measured spectra.
Both, the spatial and spectral extension of the short lived brightening at a position of $x = 2.5\,-\,3$\,arcs in the measured and at $x = 6$\,arcs in the synthetic spectrum are very similar.
In the synthetic spectrum, a brightening of about 0.5\,arcs in size appears in the line core.

Owing to the very steep change in effective formation height range with wavelength from core to wing, the fluctospheric pattern is only visible in the line core maps.
Already a few wavelength positions further into the wing, the intensity pattern is dominated by contribution from the layers below.
The imprint of reversed granulation thus already appears when observing at wavelengths close to the line center ($\Delta\lambda \approx \pm 30$\,pm.
The rather high velocities in the fluctosphere cause Doppler shifts that in principle mix the intensity signature from different layers from pixel to pixel and possibly on a sub-pixel scale.
One should thus beware that despite the narrow transmission range of the IBIS instrument the Ca\,II\,854.2\,nm line core maps could contain significant contributions from an extended height range in the atmosphere.
Nevertheless, we are confident that the intensity pattern observed here is indeed a fluctospheric shock wave pattern as seen in numerical simulations and not the reversed granulation pattern at mid-photospheric heights.
Already visual inspection of the image sequence indicates that the two-dimensional pattern evolves faster than known for reversed granulation.

This finding is quantified here in terms of the typical time scales with which the two-dimensional pattern seen at different wavelength positions in the Ca\,II infrared line evolve.
We calculate the two-dimensional auto-correlation function
\begin{eqnarray}
\mathcal{A}_{F}(\tau) &=& \frac{1}{\sum_{i_{t}=0}^{N_{T}-1}\langle \left ( F(i_{x},i_{y},i_{t})\right )^2 \rangle_{x,y}} \times \nonumber\\
&&\sum_{i_{t}=0}^{N_{T}-\tau-1} \langle \left ( F(i_{x},i_{y},i_{t}) - \bar{F} \right ) \times\\
&&\quad\quad\quad\left (F(i_{x},i_{y},i_{t}+\tau) - \bar{F} \right ) \rangle_{x,y} \nonumber
\enspace,
\end{eqnarray}
for the discrete, time-dependent function $F(x,y,t)$ sampled with $N_{T}$ time steps, which here is an intensity map sequence at a fixed wavelength.
The brackets $\langle\ \cdot\ \rangle_{x,y}$ represent the average over the spatial elements and $\bar{F} = \langle F(x,y,t) \rangle_{x,y,t}$ represents the average of all elements.
It should be noted, that two images with time lag $t_\mathrm{evol}$ are identical if $\mathcal{A}_{F}(t_\mathrm{evol})=1$ or perfectly anti-correlated for $\mathcal{A}_{F}(t_\mathrm{evol})=-1$.
In case $\mathcal{A}_{F}(t_\mathrm{evol})=0$, two images with time lag $t_\mathrm{evol}$ show no correlation at all and the scene has changed completely so that no structure related to the first image is present anymore.
The decay time scale $t_\mathrm{evol}$ of the pattern is commonly defined as the point at which $\mathcal{A}_{F}(\tau)$ has dropped to $\mathcal{A}_{F}(t_\mathrm{evol})=1/e$.
The values of $t_\mathrm{evol}$  were computed for continuum, line wing and line core intensity time sequences of the Ca II infrared line in weak-field regions.

For the Ca\,II IR line core at 854.216\,nm, the correlation time is here evaluated to $t_\mathrm{evol}^{\mbox{\tiny CaIR core}} \approx 59$\,s.
It matches the number found in previous observations of the fluctosphere using the Ca\,II K line at 393.3\,nm with a broader FWHM of 30\,pm \citep{2006A&A...459L...9W}.
In that work, a rapidly changing pattern governs the correlation time which is very similar to the pattern detected in this work in the Ca\,II IR line.
In contrast, the intensity pattern observed in the line wing of the Ca II IR line at 854.246\,nm in data set 1 clearly exhibits reversed granulation.
That pattern evolves on a scale of $t_\mathrm{evol}^{\mbox{\tiny CaIR wing}} \approx 2.5$\,min.
\citet{2005A&A...431..687L} used the observed intensity in the Ca II H line (396.883\,nm) to analyze the correlation times of the 2D structures visible.
Their observations -- obtained using an interference filter with a full width at half maximum (FWHM) of 0.1758\,nm -- show mainly the structures of reversed granulation with $t_\mathrm{evol}^{\mbox{\tiny CaH wing}} \approx 2$ minutes.
This is caused by the rather broad filter range, which includes a significant part of the line wings.
Consequently, the contributions of the brighter wings to the integrated intensity is much larger than for the comparatively dark core.
This way the reversed granulation pattern, which is formed in the middle photosphere, is the dominant pattern in the observed intensity maps.
Nevertheless, because the filter was centered on the line core of the Ca II H line, contributions from the fast changing fluctospheric layers \citep{2006A&A...459L...9W} are mixed with those from the lower layers \citep{2006ASPC..354..313U}.
This leads to shorter correlation times than those of pure reversed granulation.
Separating the faster evolving fluctospheric pattern from reversed granulation requires
filters with a very narrow transmitted wavelength range, like the FWHM of 4.6\,pm for IBIS as used here.

\begin{figure*}[t]
\centering
\includegraphics[width=\textwidth]{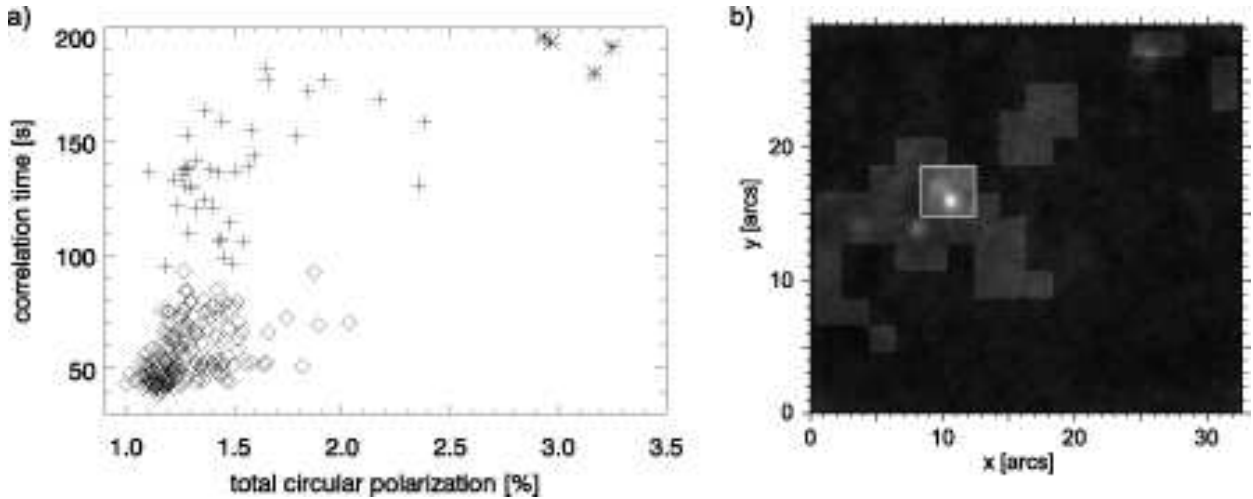}
\caption{
a) Scatter plot of correlation time (computed in subfields of 3.74$^{2}$ arcs$^{2}$ size) versus mean total circular polarization using data set~1.
b) Origin of the data points displayed as symbols in a), overlayed over the temporal average of the total circular polarization of data set~1: $\Diamond$ originate from the dark areas, $+$ originate from the gray subfields and $*$ originate from the white subfields (marked with a white box).}
\label{fig:auto}
\end{figure*}

Figure~\ref{fig:auto}~a) shows a scatter plot of evolution time scale versus total circular polarization.
The correlation function was computed in fields with 3.74$^2$\,arcs$^2$ field size to sample our FOV on a finer grid.
On average, the resultant evolution time scales were not changed significantly by the decrease of the subfield size.
We now discriminate between different types of subfields by imposing thresholds for evolution time scale and total circular polarization.
The individual subfields are correspondingly classified.
The different gray scales in Fig.~\ref{fig:auto}~b) show the distribution of the different types in the FOV.
The subfields with the longest correlation time (white patches) are connected to small-scale, long lasting structures that are likely related to network magnetic field \citep[see also][]{2006A&A...459L...9W}.

Not only vertical fields play a role in the appearance and dynamics of structures visible in the Ca II IR line, but also horizontal fields, as the gray subfields are primarily located above the structure in total linear polarization visible in Fig.~\ref{fig:mag}.
It has long been known that the magnetic flux plays an important role in emerging intensity in the chromosphere \citep{1968SoPh....3..367B}, and oscillations \citep{1993ApJ...414..345L}, as it impacts e.g. the gas pressure and thus the contribution function.
The structures with short life times prevail in regions with magnetic flux below the detection limit of IBIS.
Obviously, the magnetic field has an impact on the dynamics of the measurable intensity structures that exist in the ``mid-chromosphere''.

However, the cadence of 27\,s may not be sufficient to reliably detect time scales that are lower than twice the cadence of the data set.
A shorter cadence is certainly desirable, in particular because numerical simulations indicate time scales that are much shorter.
Such observations will become possible in the near future -- an IBIS upgrade with faster cameras is currently in progress.

\section{Conclusion}\label{sec:con}
Spectro-polarimetric observations of the solar atmosphere with a focus on
heights equivalent to the low chromosphere have been presented in this work.
We find that fluctospheric regions of the solar atmosphere, i.e. the weak field domain between the photosphere and the magnetically dominated chromosphere, display an intensity pattern that shows structures of similar spatial scales as reversed granulation but is evolving on much faster time scales.
From comparison to the predicted synthesized intensity of three-dimensional radiation hydrodynamic models of the fluctosphere, we conclude that the pattern is the intensity signature of propagating, interacting shock waves.
The pattern's dynamic time scale of 59\,s is comparable to the 20--30\,s of temperature cuts at the height of 1000\,km above $\tau_{cont}=1$, when taking into account that seeing and finite spectral passband of the instrument will lead to prolonged correlation times.

We propose a method to derive the 3D topology of the absolute LOS magnetic field flux per pixel. 
The method works for simple cases in non-emitting lines like the Ca II IR line in the Quiet Sun.
For such cases, we were capable to reconstruct the topology of a strong magnetic structure that extends from photosphere to chromosphere.
The structure becomes weaker with height, but does develop a filament like structure when recovering the field using the line core.
The method is not capable to infer filling factors, and thus cannot reproduce the field strength of an unresolved magnetic element. 

A complementary dataset with higher signal-to-noise ratio shows structures in total linear polarization encirceling strong magnetic elements. 
This strongly suggests a flux tube funnel structure in the chromosphere.
However, the location of the horizontal field remains unclear, and a static flux tube funnel appears to be a too simplified model for magnetic fields in the chromosphere.

Highly spatially resolved measurements of the magnetic fields in the chromosphere are important when trying to gather a deeper understanding of the chromospheric energy balance.
The work presented here clearly shows that, despite the usage of advanced techniques, the signal-to-noise ratio of currently available data prohibits the detection of weak field strengths and definitely not at the temporal resolution which is implied by the intensity variations seen in chromospheric/fluctospheric diagnostics such as the Ca IR line core.
Ultimately, telescopes with significantly larger apertures are needed to get a more detailed picture of the chromosphere.
The difficulties of the observation of weak magnetic fields in the chromosphere are only surpassed by the interpretation of such data.
In general, the height origins of the Stokes signals are unclear and do not only depend on contribution function but also on height distribution of the magnetic field strength.
In addition, discontinuities and steep gradients in magnetic field strength are difficult to detect and remain a fundamental problem of all methods to recover the magnetic field vector from measurements with methods available today.

\acknowledgements
SWB acknowledges support through a Marie Curie Intra-European Fellowship of the
European Commission (6th Framework Programme, FP6-2005-Mobility-5, Proposal No.
042049).
The authors would like to thank A. Tritschler for support during the data reduction.

\bibliographystyle{plainnat}

\end{document}